\documentclass[aps,prl,twocolumn,showpacs,superscriptaddress,groupedaddress]{revtex4-1}  
\usepackage[FIGTOPCAP]{subfigure}
\usepackage{graphicx}  
\usepackage{dcolumn}   
\usepackage{bm}        
\usepackage{amssymb}   
\begin{document}

\title{Deterministic control of broadband light through a multiply scattering medium via the multispectral transmission matrix} 
\author{
\large
\textsf{Daria Andreoli$^{1,2}$, Giorgio Volpe$^{1,3}$, S\'{e}bastien Popoff$^{2}$, Ori Katz$^{1,2}$, Samuel Gr\'{e}sillon$^{2}$, Sylvain Gigan$^{1}$}\\[2mm] 
\normalsize $^1$Laboratoire Kastler Brossel, ENS, CNRS UMR 8552, 24 rue Lhomond 75005 Paris\\ 
\normalsize $^2$Institut Langevin, UPMC, ESPCI ParisTech, CNRS UMR 7587, 1 rue Jussieu, 75005 Paris\\ 
\normalsize $^3$Department of Chemistry, University College London, 20 Gordon Street, London WC1H 0AJ, UK\\ 
}\vspace{5cm}
\date{\today}


\begin{abstract}
We present a method to measure the spectrally-resolved transmission matrix of a multiply scattering medium, thus allowing for the deterministic spatiospectral control of a broadband light source by means of wavefront shaping. As a demonstration, we show how the medium can be used to selectively focus one or many spectral components of a femtosecond pulse, and how it can be turned into a controllable dispersive optical element to spatially separate different spectral components to arbitrary positions.
\end{abstract}

\maketitle


\section{Introduction}
Wave propagation through a multiply scattering medium  is an important problem for different disciplines, from optics to acoustics and microwaves~\cite{Sebbah,Ishimaru,Genack}, since it finds important applications from fundamental physics to biomedical imaging~\cite{Sheng, Lihong}.  

For a monochromatic coherent input beam, the medium typically generates a speckle pattern~\cite{Goodman} - the result of complex interference between the scattered waves - that is usually considered a major drawback for imaging and focusing. In optics, wavefront shaping by spatial light modulators (SLMs) has emerged as a powerful tool to control the speckle and thus light propagation through such complex media~\cite{Fink}. Different approaches have been proposed, in particular via optimization methods~\cite{Vellekoop1}, to focus light through~\cite{Vellekoop2} or into~\cite{Vellekoop3} a complex medium. However, a more general way to understand light propagation is to describe it via its transmission matrix~(TM)~\cite{Beenakker}, {which in essence describes the fact that propagation through a disordered medium, however complex, remains a linear process, and therefore can be described by a linear operator $\mathbf{H}$ linking input to output modes, i.e. it can be described as $\mathbf{E}^{out}=\mathbf{H} \cdot \mathbf{E}^{in}$.} Recently, we introduced a method to measure the monochromatic transmission matrix of a complex medium and to subsequently demonstrate focusing and imaging through it~\cite{Popoff,Popoff2}. {Matricial methods are now widely used to characterize disordered media~\cite{Choi2011_1, Choi2011_2} or multimode fibers ~\cite{Choi2012, Psaltis, Di_Leonardo, Dholakia}, and even focusing inside complex media~\cite{Chaigne, Yaqoob}. }

Generally speaking, the medium has a spectrally dependent response, with a spectral correlation bandwidth, $\Delta\lambda_c$, that is inversely related to the confinement time of light in the medium $\Delta t$~($\Delta\lambda_c=\frac{\lambda^2}{c\Delta t}$)~\cite{Small, Beijnum, Faez, Curry}.The monochromatic treatment is, therefore, only valid as long as the bandwidth of the source is contained within this spectral correlation bandwidth. Consequently, when light from a broadband source passes through a multiply-scattering medium, its different spectral components generates different speckle patterns. This is the case where a short pulse - shorter than  the average confinement time through the medium - is scattered in space and spread in time. Thanks to the inherent coupling between spatial and spectral degrees of freedom in a scattering medium~\cite{Lemoult}, it is still possible to achieve spatiotemporal focusing through it by optimising the spatial input wavefront~\cite{Aulbach,Katz,Mertz}.

In this work, we show that we are able to measure the multispectral transmission matrix~(MSTM) of a complex medium -~i.e. the tridimensional matrix that is essentially a set of monochromatic TMs~- for several chosen wavelengths. First, we show that we are able to retrieve the spectral correlation bandwidth of the medium from the spectral correlations in the MSTM. Then, exploiting the MSTM, we can achieve efficient and deterministic spatiospectral control of a broadband and spatially coherent light source, here a femtosecond pulse, by spatial-only wavefront shaping. {The MSTM gives mores flexibility than both the optimization~\cite{Aulbach,Katz} and the pulse shaping approach~\cite{McCabe} to control separately and individually spatial and spectral components of the output fields.} We demonstrate selective spatial focusing of a given wavelength as well as multispectral focusing. Finally, we show that the medium can be turned into a controllable dispersive optical element to focus different spectral components of the pulse at different spatial positions, similarly to a grating~\cite{Freund,Park}.

\begin{figure}[htbp]
\centering
\includegraphics[width=8.1cm]{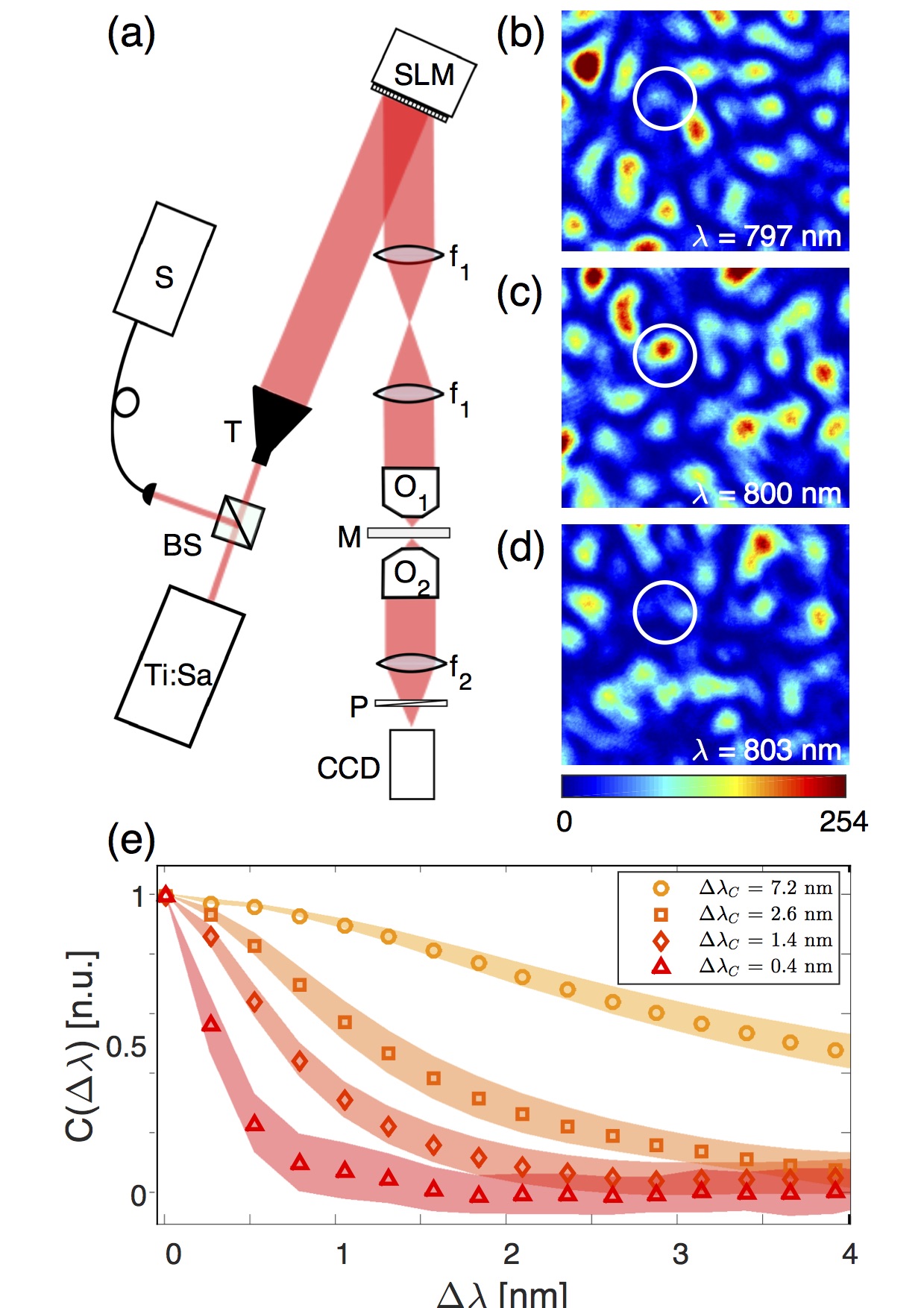}
\caption{ (a)~Experimental setup. A tunable femtosecond Ti:Sapphire laser~(Ti:Sa), working either in CW or pulsed mode, is expanded by a telescope~(T) to illuminate a two-dimensional phase-only spatial light modulator~(SLM) operated in reflection. A $4f$ system ($f_1 = 75$~mm) conjugates the SLM with the back focal plane of a low numerical aperture objective~(O$_1$, NA = 0.3) that focuses the laser beam on a strongly scattering medium~(M). The transmitted light is then imaged on a CCD camera after polarization selection by a linear polariser~(P). A spectrometer~(S) is used to monitor the output wavelength of the laser. (b-d)~Examples of the speckle patterns at the output of a ZnO sample with spectral correlation bandwidth $\Delta\lambda_c = 1.6$ nm for three totally uncorrelated wavelengths: (b)~$\lambda = 797$~nm, (c)~$\lambda = 800$~nm, and (d)~$\lambda = 803$~nm. The white circles highlight the decorrelation of one bright speckle grain at the central wavelength $\lambda = 800$~nm. (e)~Normalised spectral correlation functions $C(\Delta\lambda)$ of different strongly scattering media obtained by correlating different speckle images acquired while scanning the laser wavelength from $\lambda = 794$~nm to $\lambda = 806$~nm with a step of $\Delta\lambda = 0.2$~nm. The full width half maximum~(FWHM) of these curves gives the spectral correlation bandwidth $\Delta\lambda_c$ of the different samples: TiO$_2$ with $\Delta\lambda_c = 7.2$~nm ~(circles), ZnO with $\Delta\lambda_c = 2.6$~nm (squares), ZnO with $\Delta\lambda_c = 1.4$~nm~(diamonds), and ZnO $\Delta\lambda_c = 0.4$~nm~(triangles). For the ZnO samples the correlation bandwidth was lowered by increasing the thickness of the scattering layer. The shaded areas represent one standard deviation around the average values.}\label{fig:1}
\end{figure}

\section{Methods}

In essence, in the monochromatic TM measurement approach, a set of input wavefronts, that describes a complete basis of the input modes, is sent through the medium. Part of the beam remains unmodulated and generates a stationary speckle that serves as a reference~\cite{Popoff}. Using phase-shifting interferometry, the amplitude and phase of the output signal can then be retrieved from a set of speckle images. In order to measure the full MSTM, the laser is operated in CW mode at a given wavelength followed by the procedure for measuring the monochromatic TM of the medium~\cite{Popoff}; this step is then reiterated at every desired wavelength by changing the laser wavelength. In this 3D matrix, the output field at the $i$-th pixel on the CCD camera is connected to the input field by

\begin{equation}
\label{MSTM}
E_i^{out}(\omega_k) = \sum^{N}_{j=1}  h_{i,j,k} E_j^{in}  (\omega_k),
\end{equation}

where $ h_{i,j,k}$ is the element of the MSTM linking the $i$-th output CCD pixel to the $j$-th input SLM pixel at frequency $\omega_k$. Just as in~\cite{Popoff}, the reference is an unknown speckle whose amplitude and phase information is encoded in the $h_{i,j,k}$ coefficients. Since this reference is frequency dependent, the information on the relative local phase between monochromatic TMs cannot be accessed in this type of measurement. This measurement, therefore, precludes exact temporal control but still allows spatiospectral control of a broadband light source.

Fig.~\ref{fig:1}(a) shows a simplified scheme of the experimental setup used to measure the MSTM. A titanium-sapphire laser~(Ti:Sa, Spectra-Physics Mai-Tai) illuminates a two-dimensional phase-only spatial light modulator (SLM, Hamamatsu X10468-04). The laser is meant to work in mode-locked pulsed regime~(central wavelength tunable between $710$~nm and $920$~nm, pulse duration $\sim100$~fs and corresponding bandwidth $\Delta\lambda_L\sim10$~nm) but it can also be unlocked and operated as a continuous wave~(CW), tunable laser between $710$~nm and $920$~nm with a coherence length of $>10$~cm. With such a coherence length, the laser in CW can be considered monochromatic for the scattering media studied in this work. A spectrometer~(HR400 Ocean Optics) is used to monitor the spectrum and the mode of operation of the laser, with a spectral resolution of 0.2~nm. The illuminated SLM pixels are grouped into macro-pixels, each of which induces a controllable phase shift $\phi_j$ between $0$ and $2 \pi$. The light is reflected by the SLM and passes through a strongly scattering layer of ZnO or TiO$_2$ powder of different thicknesses~\cite{Garcia}. Figs.~\ref{fig:1}(b-d) show examples of measured speckle patterns transmitted through the same zone of one sample at different wavelengths but for a same input wavefront: these speckles are strongly wavelength-dependent. By correlating speckle images taken while tuning the wavelength of the laser in CW mode by steps of 0.2~nm, the spectral correlation bandwidth of the medium can be quantified~\cite{Beijnum}. Fig.~\ref{fig:1}(e) shows these correlation curves for samples of different thicknesses and materials. Their full-width-half-maximum defines the spectral correlation bandwidth $\Delta\lambda_c$ of the corresponding medium. 

\begin{figure}[htbp]
\centering
\includegraphics[width=8.1cm]{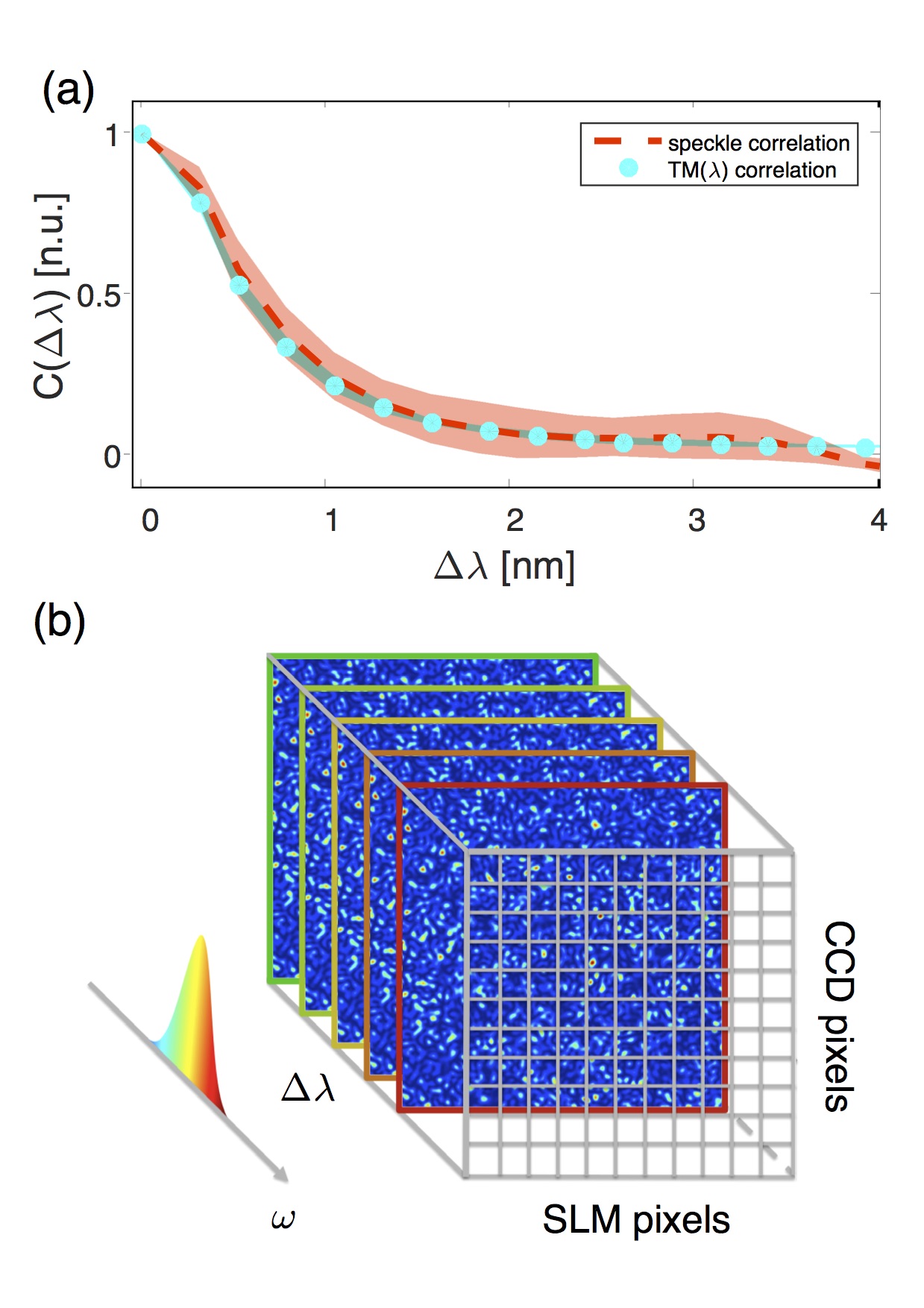}
\caption{Measurement of the multispectral transmission matrix~(MSTM). (a)~The cyan dots show the spectral correlation $C(\Delta\lambda)$ among different monochromatic transmission matrices~(TMs) for a ZnO medium~($\Delta\lambda_c = 1.6$~nm) measured in the wavelength range between $\lambda = 794$~nm and $\lambda = 806$~nm with a step of $\Delta\lambda = 0.2$~nm. This correlation curve is consistent with the spectral correlation curve $C(\Delta\lambda)$ calculated by directly correlating speckle intensity images~(Fig.\ref{fig:1}(e)) and here represented by the red dashed line. The shaded areas represent one standard deviation around the average values. (b)~In order to maximise the information about the scattering process and minimise the measurement time, the MSTM is a 3D matrix composed of $N$ monochromatic matrices recorded at regular intervals within the bandwidth $\Delta\lambda_L$ of the polychromatic pulse propagating though the scattering medium.}\label{fig:2}
\end{figure}

Fig.~\ref{fig:2}(a) shows that the correlation between different spectral components of the MSTM is consistent with the spectral correlation bandwidth of the medium as measured in Fig.~\ref{fig:1}(e). 
In order to maximise the information and minimise the amount of measurements required to control a polychromatic pulse of bandwidth $\Delta \lambda_L$, the spectral sampling of the MSTM can be chosen to match the spectral correlation bandwidth $\Delta\lambda_c$ of the medium. The elements of the MSTM are sufficiently uncorrelated and yet contain enough information to describe accurately the scattering process. Here, therefore, the medium response is sampled by a MSTM composed of $N_{\lambda}$ monochromatic TMs with $N_{\lambda} = \frac{\Delta\lambda_L}{\Delta\lambda_c}$ (Fig.~\ref{fig:2}(b)).

In the following, all measurements are performed on a sample with spectral correlation bandwidth of 1.6~nm, corresponding to an average confinement time of 625~fs. The input light is a broadband pulse with a bandwidth of~$\sim10$~nm, corresponding to a Fourier-limited duration of~$\sim100$~fs. Therefore, an experimental $N_{\lambda}=6$ is chosen. The number of input macro-pixels is set to 256, corresponding to $16 \times 16$ macro-pixels, and, on the CCD camera, an observation region, approximately corresponding to $10 \times 10$ speckle grains, is selected. 

\section{Results and discussion}

\subsection{Spatial focusing}

\begin{figure}[htbp]
\centering
\includegraphics[width=8.1cm]{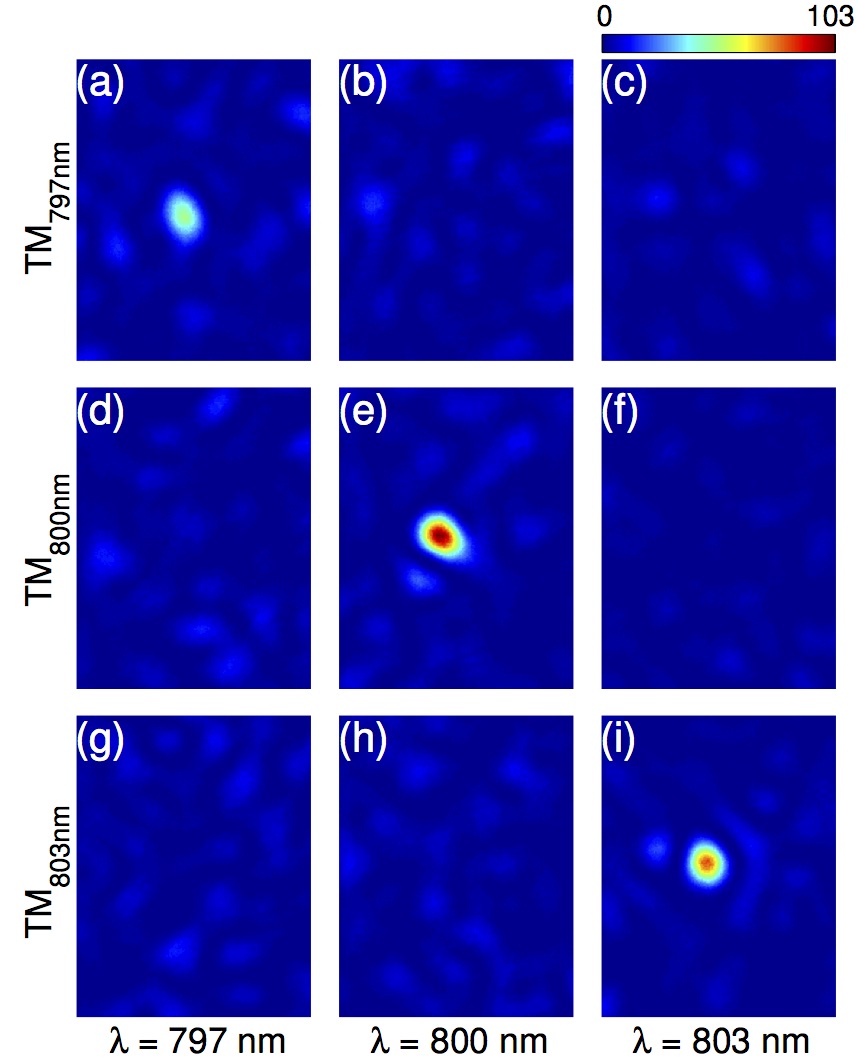}
\caption{Selective monochromatic focusing through a strongly scattering medium with the MSTM. (a-i)~Focusing images of a monochromatic beam at (a,d,g) $\lambda = 797$~nm, (b,e,h) $\lambda = 800$~nm, and (c,f,i)~$\lambda = 803$~nm, using the phase mask given by a vertical section of the MSTM, in the CCD-SLM pixels plane,~(Fig.\ref{fig:2}(b)) recorded at (a-c) $\lambda = 797$~nm,~(d-f) $\lambda = 800$~nm, and~(g-i) $\lambda = 803$~nm. A focus appears only when the TM recorded at the corresponding wavelength of the beam is selected in (a) for $\lambda = 797$~nm, in (e) for $\lambda = 800$~nm, and in (i) for $\lambda = 803$~nm, respectively.} \label{fig:3}
\end{figure}

The first most straightforward use of the MSTM is as a set of monochromatic TMs to focus light at a specific wavelength through a strongly scattering medium by phase conjugation, as shown in Fig.~\ref{fig:3}. This corresponds to selecting a vertical section of the MSTM~(in the CCD-SLM pixels plane) in Fig.~\ref{fig:2}(b). In this case, only the fraction of the input beam at the chosen wavelength~(within the spectral correlation bandwidth) will be effectively controlled~\cite{Beijnum}. Figs.~\ref{fig:3}(a-i) show focusing images of monochromatic waves at various wavelengths~(columns) using the phase mask given by a vertical section of the MSTM recorded at a different wavelength~(rows). As expected, a focus appears only when the incident wavelength matches the wavelength at which the monochromatic TM was recorded. 

\subsection{Multispectral focusing}

\begin{figure}[htbp]
\centering
\includegraphics[width=8.1cm]{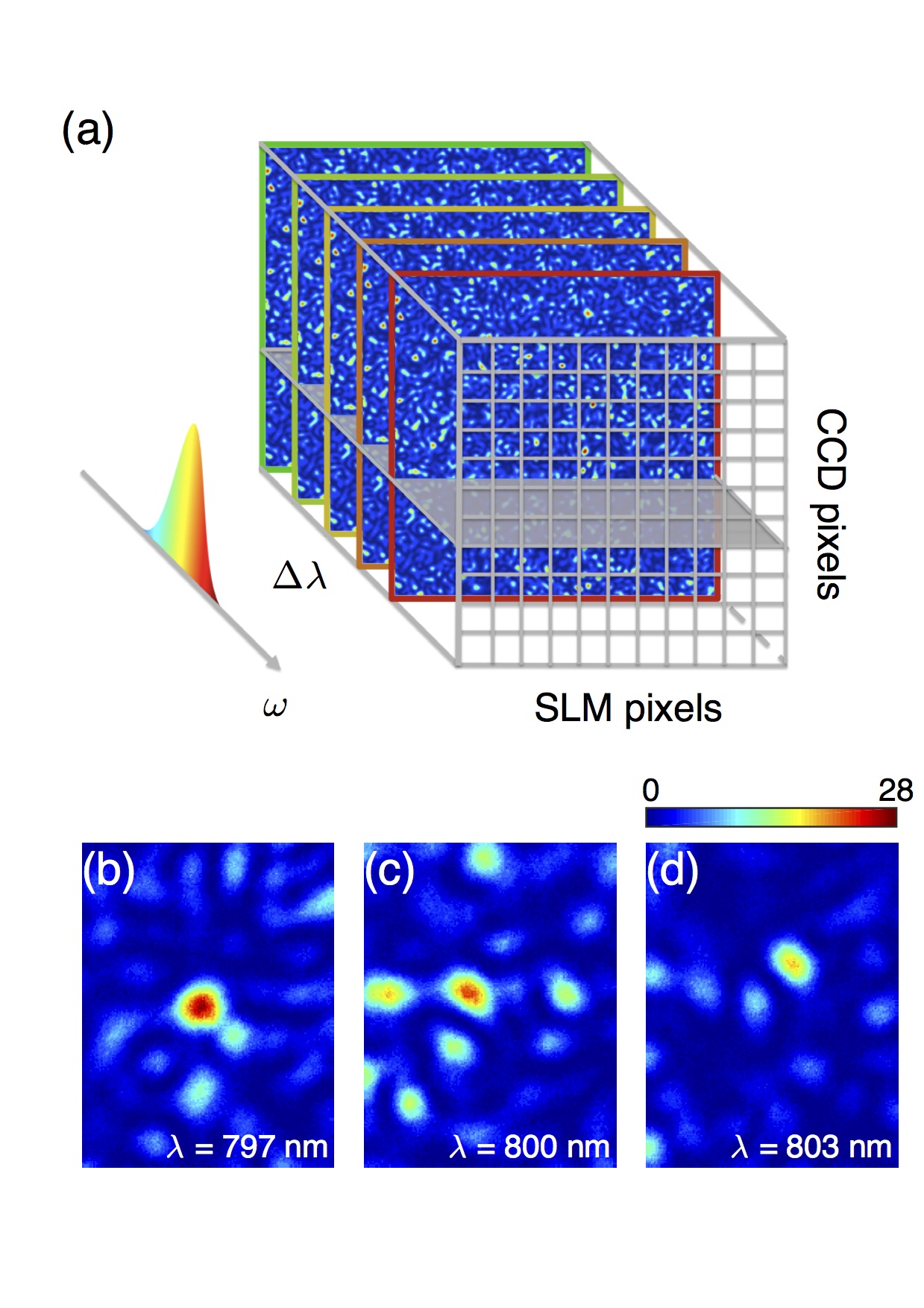}
\caption{Simultaneous focusing of different wavelengths at the same position (a) A transversal cut of the MSTM~(grey area) allows one to spatially focus on a given CCD pixel all the independent wavelengths within the bandwidth $\Delta\lambda_L$.  (b-d)~Focusing images  for a monochromatic beam at various frequency for this optimal phase mask: (b) $\lambda = 797$~nm, (c) for $\lambda = 800$~nm, and (d) $\lambda = 803$~nm. The focusing efficiency is reduced by a factor roughly $N_\lambda$ compared to Fig~\ref{fig:3}(a)(e) and (i) since the phase mask tries to focus many wavelengths simultaneously.}\label{fig:4}
\end{figure}

Just like it is possible to focus in the monochromatic case in several points simultaneously by adding coherently the corresponding phase masks on the SLM~\cite{Popoff}, it is also possible to focus several wavelengths at the same position by displaying a coherent sum of different phase-conjugate wavefronts for different wavelengths extracted from the MSTM. When one considers a single output position, an elegant way to formalize the above solution is to take a horizontal section of the MSTM as shown in Fig.~\ref{fig:4}(a). This section is a two-dimensional matrix $\tilde{\mathbf{H}}$ that links the spectral components of the input SLM pixels to the spectral components at a given output spatial position $i_t$. In a quasi-monochromatic case, when the bandwidth of the polychromatic source is much smaller than the central wavelength, the dependency in $\omega_k$ at the input can be neglected. In this configuration, Eq.~\ref{MSTM} reduces to

\begin{equation}
\label{MSTM_transversal_simplified}
E_{i_t}^{out}(\omega_k) = \sum^{N}_{j=1}  \tilde{h}_{i_t,j,k} E_j^{in}.
\end{equation}

This problem is equivalent to the one treated in~\cite{Popoff} but now the output dimension is spectral rather than spatial. For a given desired output spectrum vector $\bf{E^{target}}$, the input vector can be obtained by phase conjugation as $\mathbf{E}^{in}=\tilde{\mathbf{H}}^\dagger \cdot \mathbf{E}^{target}$. 

{In this case, with homogeneous illumination and phase-only control, the phase of the calculated $\mathbf{E}^{in}$ needs to be displayed on the SLM. In order to focus spatially one spectral component $\omega_k$, $E^{target}(\omega_k)$ is set to 1 while all the other components are set to zero: the resulting wavefront is identical to what is obtained by phase conjugation of the monochromatic TM at $\omega_k$. In order to focus all spectral components with the same weight, the imposed condition is $\mathbf{E}^{target} = \mathbf{1}$. Using a phase masks that focuses several wavelengths simultaneously, and scanning the wavelength in CW mode, Figs.~\ref{fig:4}(b-d) show that all wavelengths are indeed focused at the same spatial position.} 

\begin{figure}[htbp]
\centering
\includegraphics[width=8.1cm]{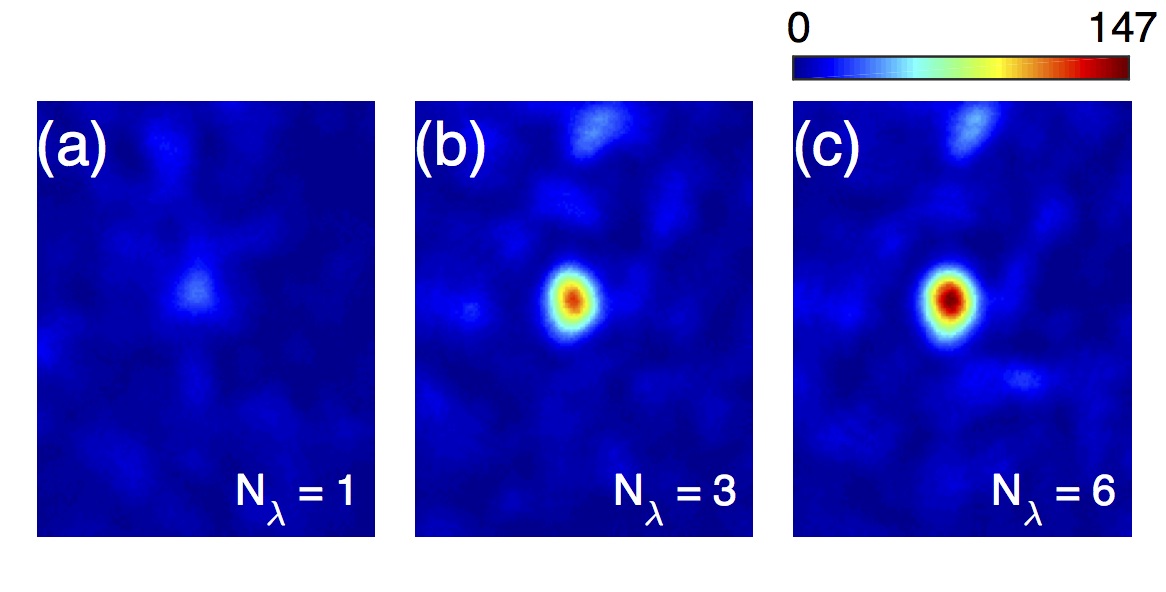}
\caption{Polychromatic focusing through a strongly scattering medium with the MSTM. (a-c)~The contrast of the spatial focus increases with $N_{\lambda}$, number of controlled wavelength: (a) SBR= 7.2 for $N_{\lambda} = 1$, (b) SBR = 26.1 for $N_{\lambda} = 3$, and (c) SBR = 30.7 for $N_{\lambda} = 6$. }\label{fig:5}
\end{figure}

{This approach becomes even more relevant for broadband illumination. Fig.~\ref{fig:5} shows, for an ultrashort pulse, how the intensity at the focal point increases over the background when the number $N_{\lambda}$ of different spectral components set to one in $\mathbf{E}^{target}$, all other components being set to zero. In Fig.~\ref{fig:5}(c), all 6 independent spectral components of the broadband pulse measured with the MSTM are controlled, showing clear spatial focusing. The signal to backroung ratio~(SBR), defined as the ratio between the average value of all the pixels inside the FWHM and the average value of the speckle image before focusing, is~30.7. Since these are linear images, they are not sensitive to possible fluctuations of the temporal duration of the pulse at this position. When we use the speckle as reference,  the relative phase between the different spectral components is unknown, and no temporal compression of the pulse occurs.}

\begin{figure*}[htbp]
\centering
\includegraphics[width=13cm]{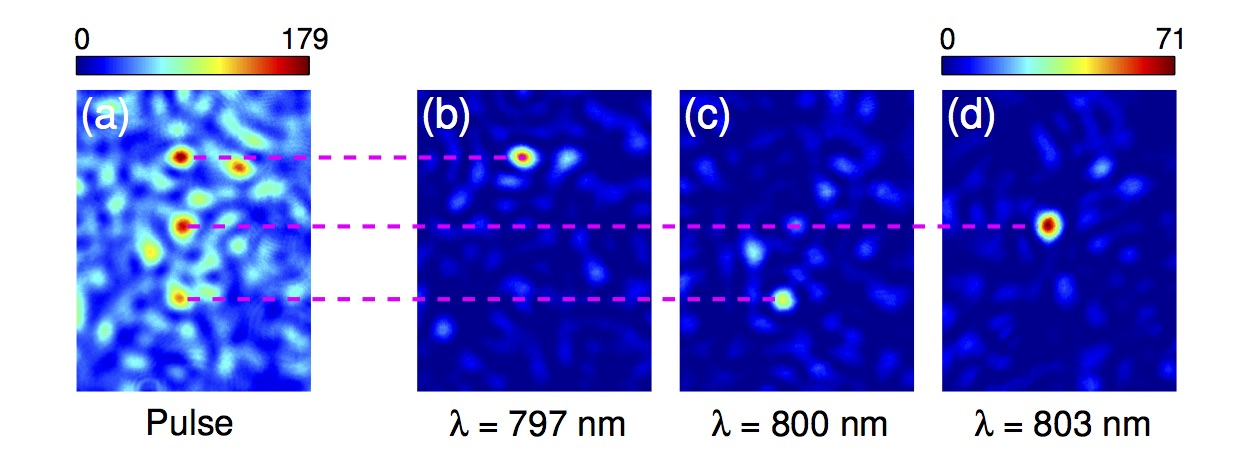}
\caption{Implementation of a generalized grating through a strongly scattering medium with the MSTM. (a) Simultaneous spatial focusing in different output points of three independent wavelengths ($\lambda = 797$~nm, $\lambda = 800$~nm, and $\lambda = 803$~nm) within the bandwidth $\Delta\lambda_L$ of a polychromatic pulse using the MSTM. (b-d) Focusing images of a monochromatic beam at (b) $\lambda = 797$~nm, (c) $\lambda = 800$ nm, and (d) $\lambda = 803$~nm, using the same phase mask of (a). The three beams are focused in independent foci corresponding to the three output points in (a), as highlighted by the dashed lines.}\label{fig:6}
\end{figure*}

\subsection{Turning the medium into a generalized grating}

As a last example, Fig.~\ref{fig:6} demonstrates another possible use of the MSTM by focusing different wavelengths to different positions. In this way the scattering medium is used as a deterministic dispersive optical element~\cite{Freund}. In Fig.~\ref{fig:6}(a), three different wavelengths of a broadband pulse are focused at three arbitrary positions along a line. Scanning the wavelength of a monochromatic laser reveals that each focus corresponds indeed to a different wavelength within the pulse (797~nm in Fig.~\ref{fig:6}(b), 800~nm in Fig.~\ref{fig:6}(c) and 803~nm in Fig.~\ref{fig:6}(d)). These results can be interpreted as a generalized grating. {In terms of resolution and efficiency, the main difference between our approach and a conventional dispersive element is that, in our case, scattering by the medium itself determines both spectral resolution and spatial localization. Spectral resolution is given by the spectral correlation bandwidth of the medium, linked to the confinement time of the light in the medium, whereas the latter is diffraction-limited, i.e. limited by the effective numerical aperture of the medium. On the other hand, the efficiency of the spatio-spectral focusing, given by the SBR, is ultimately limited by how many degrees of freedom are controlled, i.e. how many SLM pixels and how many speckle grains are controlled simultaneously. In practice, the limitation in number of degrees of control is due to the measurement time, which is limited by the stability of the experimental setup and the medium. Here the measurement time is of the order of 15 minutes for 256 input modes and 6 wavelengths. This can easily be scaled up by increasing the stability and the modulation speed.} The spectral control is, in this case, deterministic~\cite{Cao}. Interestingly, unlike a grating, the spatial dispersion of the different spectral foci can be arbitrary as demonstrated in Fig.~\ref{fig:6}, since each independent wavelength can be focused at any spatial position.  

\section{Conclusion}

In conclusion, we introduced a method  to measure the spatially and spectrally resolved transmission matrix, or  MSTM, of a multiply scattering medium. Compared to other approaches based on optimization, the main advantage of our technique is that it is deterministic. Exploiting the information contained in the MSTM, we are able to selectively focus one or many spectral components of a broadband light source to one or many arbitrary spatial positions. As a proof of principle, we demonstrate multispectral focusing of a short pulse, and show the medium can be turned into a deterministic dispersive optical element, thus generalizing the concept of grating. Here we demonstrate spatiospectral control, and our results are valid not only for short laser pulse, but also for temporally incoherent sources such as SLEDs or supercontinuum sources. Further control of the spectral phase with a plane wave as reference should enable deterministic spatiotemporal control of an ultrashort pulse. 


\section{Acknowledgements}
This work was funded by the European Research Council (grant no. 278025), ANR ROCOCO and Programme Emergence(s) from the city of Paris. O.K. was supported by the Marie Curie Intra-European fellowship for career development (IEF). We thank David Martina and Mickael Mounaix for helpful discussions. 

\section{Author Contributions Statement}
S.G. (Gigan) conceived the project. D.A., S.G. and S.G. planned the experiment. D.A. performed the experiment and analyzed the data. S.P. proposed the analytic formalism for spatiospectral control. D.A., G.V. and S.G. (Gigan) wrote the manuscript.  All authors discussed the results and  revised the manuscript.


\begin{thebibliography}{99} 

\bibitem[1]{Sebbah}
P. Sebbah, \emph{Waves and imaging through complex media} (Springer, 2001)
\newblock

\bibitem[2]{Ishimaru}
A. Ishimaru, \emph{Wave propagation and scattering in random media} (Wiley-IEEE Press, 1999)
\newblock

\bibitem[3]{Genack}
M. Stoytchev and A. Z. Genack, Phys. Rev. B, \textbf{55}, 14, (1997). 
\newblock

\bibitem[4]{Sheng}
P. Sheng,  \emph{Introduction to Wave Scattering, Localization, and Mesoscopic Phenoma} (Academic Press, New York, 1995).
\newblock

\bibitem[5]{Lihong}
Lihong V. Wang and Hsin-i Wu,  \emph{Biomedical Optics: Principles and Imaging} (Wiley, 2007).
\newblock 

\bibitem[6]{Goodman}
J. Goodman, \textit{Speckle Phenomena in Optics} (Roberts and Company Publishers, 2010).
\newblock

\bibitem[7]{Fink}
A. P. Mosk, A. Lagendijk, G. Lerosey and Mathias Fink, Nat. Phot. \textbf{6}, 283–292 (2012).
\newblock 

\bibitem[8]{Vellekoop1}
I. Vellekoop and A. Mosk, Optics Letters, \textbf{32}, 12 (2007).
\newblock

\bibitem[9]{Vellekoop2}
I. Vellekoop, A. Lagendijk and A. Mosk, Nature Photonics, \textbf{4}, 320-322 (2010).
\newblock

\bibitem[10]{Vellekoop3}
I. Vellekoop, E. Van Putten, A. Lagendijk and A. Mosk, Optics Express, \textbf{16}, 67-80 (2008).
\newblock

\bibitem[11]{Beenakker}
 C. W. J. Beenakker, Rev. Mod. Phys. \textbf{69}, 731 (1997).
\newblock
 
\bibitem[12]{Popoff}
S. M. Popoff,  G. Lerosey, R. Carminati, M. Fink, A. C. Boccara, and S. Gigan, Phys. Rev. Lett. \textbf{104}, 100601 (2010).
\newblock

\bibitem[13]{Popoff2}
S. M. Popoff,  G. Lerosey, M. Fink, A. C. Boccara, and S. Gigan, Nat. Commun. \textbf{1}, 81 (2010).
\newblock

\bibitem[14]{Choi2011_1}
Y Choi, TD Yang, C Fang-Yen, P Kang, KJ Lee, RR Dasari, M. S. Feld and W. Choi, Phys. Rev. Lett. \textbf{107}, 2, 023902 (2011).
\newblock

\bibitem[15]{Choi2011_2}
M Kim, Y Choi, C Yoon, W Choi, J Kim, QH Park and W Choi, Nature Photonics \textbf{6}, 581-585 (2011).
\newblock

\bibitem[16]{Dholakia}
T. \^{C}i\^{z}m\'{a}r and K. Dholakia, Optics Express, \textbf{19}, 18871-18884 (2011). 
\newblock

\bibitem[17]{Choi2012}
C Yoon, Y Choi, M Kim, J Moon, D Kim and W Choi, Opt. Lett. \textbf{37}, 21 (2012).
\newblock

\bibitem[18]{Psaltis}
I. N. Papadopoulos, S. Farahi, C. Moser and D. Psaltis, Biomed Opt Express. \textbf{4}, 260–270 (2013).
\newblock

\bibitem[19]{Di_Leonardo}
S. Bianchi, V. P. Rajamanickam, L. Ferrara, E. Di Fabrizio, C. Liberale, and R. Di Leonardo, Optics Letters, \textbf{38}, 4935-4938 (2013).
\newblock

\bibitem[20]{Chaigne}
T. Chaigne,	O. Katz,	A. C. Boccara,	M. Fink,	E. Bossy and S. Gigan, Nature Photonics \textbf{8}, 58–64 (2014).
\newblock

\bibitem[21]{Yaqoob}
Y. Choi, T. R. Hillman, W. Choi, N. Lue, R. R. Dasari, P. T. C. So, W. Choi and Z. Yaqoob, Physical Chemistry Chemical Physics \textbf{16}, 27074-27077 (2014)
\newblock

\bibitem[22]{Small}
E. Small, O. Katz, Y. Guan, and Y. Silberberg, Opt. Lett., \textbf{37}, 16 (2012)  
\newblock

\bibitem[23]{Beijnum}
F. van Beijnum, E. G. van Putten, A. Lagendijk, and A. P. Mosk, Opt. Lett., \textbf{36}, 3 (2011) 
\newblock

\bibitem[24]{Faez}
S. Faez, P. M. Johnson, and Ad Lagendijk, Phys. Rev. Lett., \textbf{103}, 053903 (2009) 
\newblock

\bibitem[25]{Curry}
N. Curry, P Bondareff, M. Leclercq, N. F. van Hulst, R. Sapienza, S. Gigan, and S. Gr\'{e}́sillon, Opt. Lett.  \textbf{36}, 17 (2011).
\newblock

\bibitem[26]{Lemoult}
F. Lemoult, G. Lerosey, J. de Rosny, and M. Fink, Phys. Rev. Lett., \textbf{103}, 173902 (2009).
\newblock

\bibitem[27]{Aulbach}
J. Aulbach, B. Gjonaj, P.Johnson, A. Mosk, and A. Lagendijk, Phys. Rev. Lett., \textbf{106}, 103901 (2011).
\newblock

\bibitem[28]{Katz}
O. Katz, E. Small, Y. Bromberg, and Y. Silberberg, Nature Photon., \textbf{5}, 372 (2011).
\newblock

\bibitem[29]{Mertz}
H. P. Paudel, C. Stockbridge, J. Mertz and T. Bifano, Optics Express, \textbf{21}, 14 (2013).
\newblock

\bibitem[30]{McCabe}
D. J. McCabe, A. Tajalli, D. R. Austin, P. Bondareff, I. A. Walmsley, S. Gigan, and B. Chatel, Nat. Commun. \textbf{2}, 447 (2011).
\newblock

\bibitem[31]{Freund}
I. Freund, Phys. A, \textbf{168}, 49–65 (1990). 
\newblock

\bibitem[32]{Park}
J. H. Park, C. Park, H. Yu, Y. H. Cho and Y. K. Park, Opt. Lett., \textbf{37}, 15 (2012). 
\newblock

\bibitem[33]{Garcia}
P. Garc\'{i}a, R. Sapienza, \'{A}. Blanco, and C. L\'{o}pez, Adv. Mater. \textbf{19}, 2597 (2007).
\newblock

\bibitem[34]{Cao}
B. Redding, S. F. Liew, R. Sarma and H. Cao, Nature Photon., \textbf{7}, 746–751 (2013).
\newblock
\end{thebibliography}
\end{document}